\documentclass[final]
	      {aipproc}
	      \layoutstyle{6x9}
	      \begin{document}
	      \title{Non Photonic $\mathrm{e-D^0}$ correlations in $\mathrm{p+p}$ and $\mathrm{Au+Au}$ collisions at $\mathrm{\sqrt{s_{NN}}=200~GeV}$}
              \author{Artemios Geromitsos for the STAR Collaboration } {
	        address={SUBATECH, Ecole de Mines, 4 rue Alfred Kastler, BP 20722,
		  44307, Nantes, France.} }

              \newcommand{\NUCL}{Nuclear Physics }
              \newcommand{\PHYSREV}{Phys. Rev.}
              \newcommand{\PHYSREVLET}{Phys. Rev. Lett.}
              \newcommand{\PHYSLET}{Phys. Lett.}
              \newcommand{\JPHYS}{J.Phys}

	      \begin{abstract}
		The sum of charm and beauty in Au+Au collisions at 200\,GeV measured through nonphotonic electrons, show
		similar suppression at high p$_T$ as light hadrons, in contrast to expectations based on the dead cone effect.
		To understand this observation, it is important to separate the charm and beauty components.
		 Non-photonic electron-$\mathrm{D^0}$ and electron-hadron azimuthal angular correlations 
		are used to disentangle the contributions from charm and beauty decays.
		 The beauty contribution  in p+p collisions at 200 GeV
		  is found to be comparable to charm  at p$_T$ $\sim$ 5.5 GeV, indicating that
		beauty may contribute significantly to the non photonic electrons from heavy flavour
		decays in Au+Au data at high p$_T$.
		Furthermore, in Au+Au collisions we
		present the status of D0 meson reconstruction using microvertexing techniques
		made possible with the addition of the silicon detectors.

	      \end{abstract}
	      \maketitle

              \section{Introduction}
		One of the most important discoveries at RHIC is the quenching of jets traversing the hot and dense 
		matter         
		created in heavy ion collisions. Jet quenching allows for an estimate of the gluon
		density of the medium. 
              One of the current puzzles at RHIC is the suppression of heavy flavors measured
	      through non photonic electrons, which is found to be similar to that observed for charged hadrons
		\cite{nonphotonic}
	      in contrast to theoretical expectations based on the dead cone effect 
		\cite{deadcone}.
              Therefore, it is of interest to determine separately the relative contributions from charm and beauty.
	      STAR uses non-photonic electron-$\mathrm{D^0}$ and electron-hadron azimuthal angular correlations 
	      in order  to disentangle the contributions from charm and bottom decays.

              \begin{figure}[htbp]
		\includegraphics[scale=0.6,angle=90]{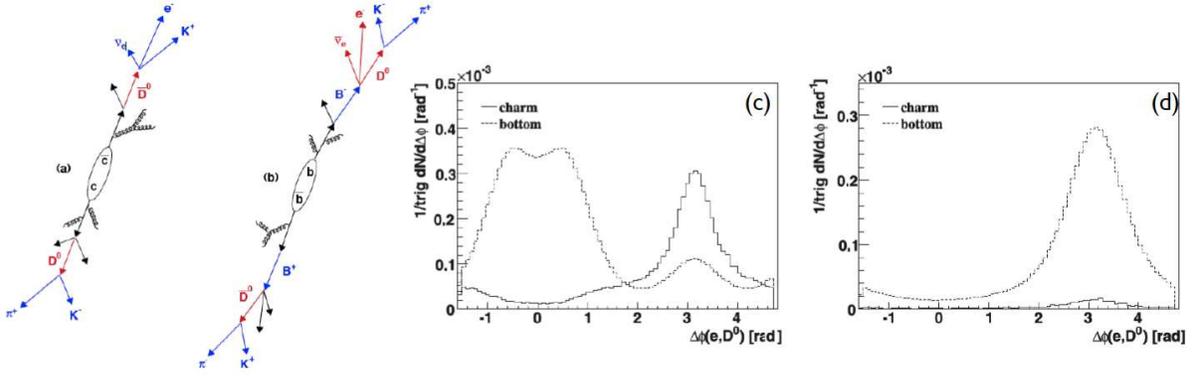}
	        \caption{Left: $c-\overline{c}$ and $b-\overline{b}$
		 decay channels considered for the e-$D^0$ study. 
		e-$D^0$ azimuthal correlations for  the like-sign e-K (c) and the unlike-sign e-K (d).}
		\label{fig:cc_n_bb_n_delta_phi}    
              \end{figure}

              \section{Analysis  and results}
		
		One method to disengtangle charm and beauty is the measurement of azimuthal angular correlations between
		non-photonic electrons and charged hadrons in p+p collisions 
		\cite{e-h}.
		The width of the near-side correlation function is expected to be larger for the semileptonic decay of beauty compared 
		to charm, due to the larger mass of the b quark.
		 The distributions for charm and beauty meson decays have been obtained from PYTHIA and 
		then simultaneously fitted to the experimental correlation function to obtain the relative
		contribution of bottom decay as a function of p$_T$ 
		up to p$_T$ of 9 GeV \cite{e-h}.

		A second method concerns  e-$D^0$ correlations resulting from $c-\overline{c}$ and $b-\overline{b}$
		 production and decay as 
		shown in fig. 1 (a) and (b)
		\cite{andre_mischke}. 
		The branching fraction of charm and beauty decaying into electrons
		is about 10\%. The electrons from these decays can be selected by the online trigger through their energy
		deposition in the STAR electromagnetic calorimeter, while the hadronic decay of the
		$D^0 \rightarrow K^- \pi^+$ can be reconstructed through the $K^- \pi^+$ invariant mass.

		Figure 1 illustrates the azimuthal angular correlation of electron and $D^0$
		for the unlike-sign electron-kaon (fig. 1, d), and the like sign electron-kaon (fig. 1, c), from a PYTHIA simulation 
		\cite{andre_mischke}.
		It is shown that the near-side peak of the like-sign e-K case is dominated by $D^0$ from beauty, while
		the away-side peak of the like-sign e-K case is dominated by $D^0$ from charm decay (fig. 1, c).
		The away-side peak of the unlike-sign e-K case is dominated by   $D^0$ from beauty (fig. 1, d).
		Therefore, the azimuthal correlation of e-$D^0$  in combination with the charge-sign requirement allows
		the separation of charm and beauty components.


              \begin{figure}[htbp]
		 \includegraphics[scale=0.3,angle=90]{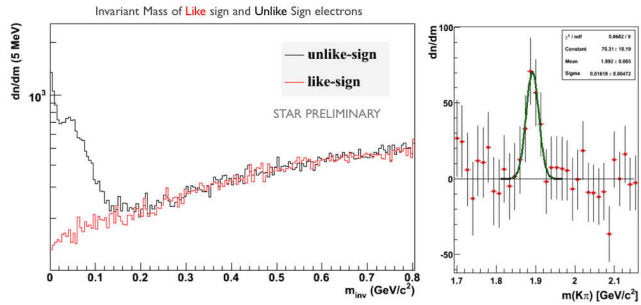}
                \caption{\textit{Left}: 
		Invariant mass of the like sign ( \textit{red}) and the unlike sign  (\textit{black}) electron pairs.
		 \textit{Right}: Invariant mass of $K \pi$ in p+p collisions at $\sqrt{s}$=200 GeV after background subtraction.}
                \label{fig:eplus_emin}
	       \end{figure}

		Electron identification uses the combined information from the Time Projection Chamber (TPC) 
		the Barrel Electromagnetic calorimeter (BEMC) and the Shower Maximum Detector (SMD).
 The STAR TPC  and the BEMC cover midrapidity ($|\eta|<1$) and full azimuthal angle.
		Electrons are identified by applying a momentum dependent cut on the ionization energy loss within
	the range  $\mathrm{3.5<dE/dx<5.0 (keV/cm)}$, a cut in the shower profile size, and by requiring the
		(TPC) momentum to (BEMC) energy ratio to be  $\mathrm{0<\frac{p}{E}<2}$.
		The electrons have contributions from the signal heavy-flavour decays, while the main background is 
		electrons from photon conversions, $\pi^0$ and $\eta$ Dalitz decays.
		 To reject the 'photonic  electrons'  the invariant mass of all e+ and e- candidates in each event is formed
		  (figure~\ref{fig:eplus_emin}, left). The like-sign 
              electron (positron) candidate (\textit{red}) is superimposed to the unlike-sign pair (\textit{black}) invariant mass.
              The crossing point at $\mathrm{150~MeV/c^2}$ indicates the corresponding cut choosen, 
		excluding the electron candidates that form invariant mass below this value.
	The partner finding efficiency was estimated to be ~$50\%$ at low $p_T$ and increases to ~$80\%$ at high $p_T$.
	
		$D^0$ mesons are reconstructed through their hadronic decay $D^0 \rightarrow K^- \pi^+$ (BR=3.89\%)
		 (fig.~\ref{fig:eplus_emin}, right).
		Kaons are required to have a dE/dx within  $\pm$ 3 $\sigma$ from the expected kaon dE/dx.
              The combinatorial background is being calculated by taking into account 
              the like sign pairs invariant mass yield $\mathrm{\sqrt{(K^{+}\pi^{+}) * (K^{-}\pi^{-}}) }$.

              \begin{figure}[htbp]
		 \includegraphics[scale=0.7,angle=0]{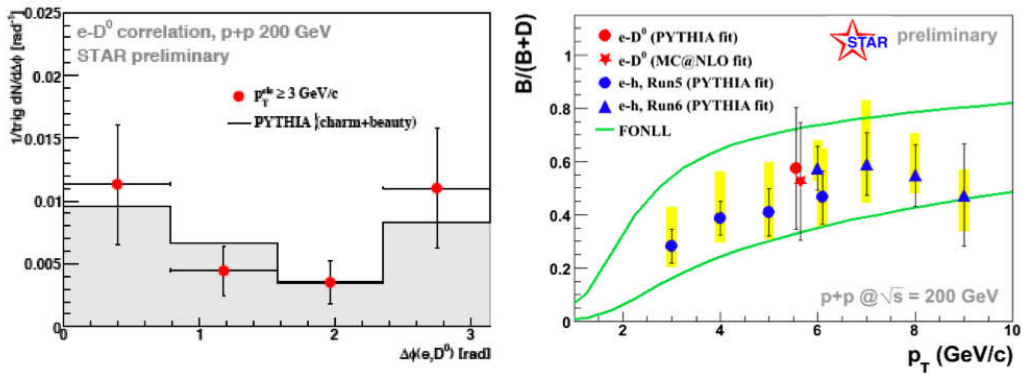}
                \caption{\textit{Left}: e-$D^0$ azimuthal angular correlations in p+p collisions at 200 GeV
		(for the like-sign e-K pairs) compared to PYTHIA, scaled by a factor 2.86.
		 \textit{Right}: Ratio of beauty to the sum of beauty and charm as a 
		function of p$_T$ in p+p collisions at 200 GeV, compared to FONLL calculations.
		In this analysis no microvertexing technique was used.}
                \label{fig:b_ratio}
	       \end{figure}

		Measurements in $p+p$ collisions without micro-vertexing techniques provide a 
		baseline from which to search in Au+Au collisions.  Figure~\ref{fig:b_ratio}
		left, shows the azimuthal correlation distribution of electrons and $D^{0}$
		mesons in $p+p$ collisions,
		 which exhibit a near- and a away- side correlation peak with
		similar yields and is compared to a PYTHIA simulation \cite{andre_mischke}.

		The relative contribution of beauty to the sum of beauty and charm 
		estimated by the methods of $e-h$ and $e-D^0$ azimuthal correlations 
		is shown in figure~\ref{fig:b_ratio} \cite{e-h}.
 		The contribution of beauty increases with  p$_T$
		and becomes 50\% at p$_T$ around 5.5 GeV.
		The data are found to be compatible with FONLL estimates.

              \begin{figure}[htbp]
                \includegraphics[scale=0.5,angle=90,trim= 0mm 0cm 0mm 0cm]{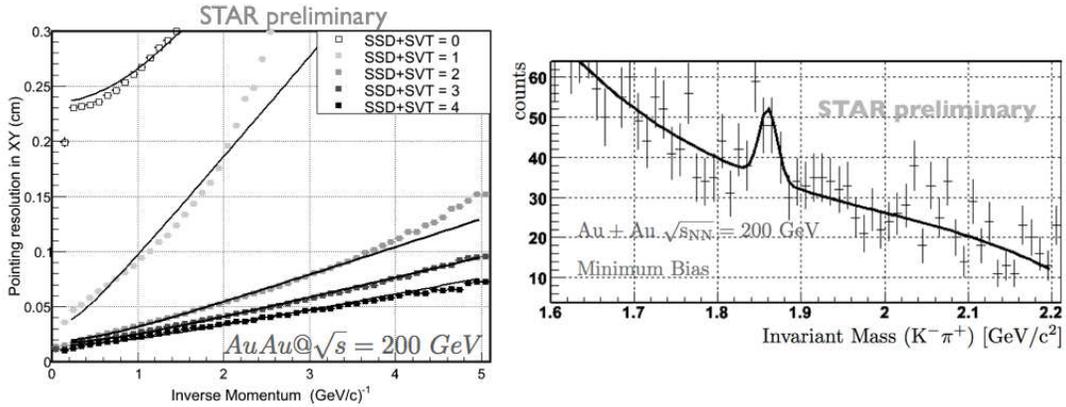}
                \caption{\textit{Left}: Transverse DCA resolution vs $1/p$. \textit{Right}:  Invariant Mass for the 
		channel $D^0 \rightarrow K^{-} \pi^{+}$ in Minimum Bias $\textrm{Au+Au}$ collisions at $\sqrt{s_{NN}}=200~GeV$. 
		In this analysis, we take advantage of the improved momentum resolution due to the silicon detectors of STAR.}
                \label{fig:minv_d0_auau}    
              \end{figure}

	 	In the analysis of Au+Au collisions at 200 GeV we are employing microvertexing techniques, 
	        taking advantage of the STAR silicon detectors to improve the S/B ratio
		for the $D^0$ inv. mass.
		The Silicon tracker consist of 3 layers of silicon drift detector (SVT) at 6.85, 10.8 and 14.5 cm  and 
		one layer of silicon strip detector (SSD) at 23 cm from the beam. 
		The silicon inner tracker of STAR leads to an improvement by an order of magnitude of the resolution
                of the Distance to Closest Approach (DCA) of charged particles to the Primary Vertex for momentum ~1 GeV
		(figure~\ref{fig:minv_d0_auau}, left) 
		\cite{yuri}.

		As a check of the micro-vertexing technique, figure~\ref{fig:minv_d0_auau}, right, shows a preliminary
		invariant mass plot $K^- \pi^+$ from peripheral Au+Au collisions at 200 GeV.
		There is a hint of a signal peak in the $D^0$ mass region, after requiring at least one hit in the Silicon inner tracker
		of STAR and a decay length less than 700 $\mu m$ among other cuts.

                \section{Summary and conclusion}
		Heavy quarks measured through non photonic electron yields show a larger suppression
		than expected. To understand this puzzle and understand better the flavour dependence
		of jet quenching, separation of charm and beauty contributions is important.
		STAR is using two methods for this purpose, electron-hadron and electron-$D^0$ azimuthal
		angular distributions.
		Using these methods it is found that the beauty contribution increases with p$_T$ and 
		becomes comparable to the charm contribution around p$_T$~5.5 GeV in p+p collisions
		at 200 GeV. 
The beauty  contribution  is found to be compatible to FONLL calculations
		  within the uncertainties.
		The analysis of $D^0$ identification in Au+Au collisions at 200 GeV
		using microvertexing techniques based on the STAR silicon inner tracker
		is underway.
		 In the near future the full TOF detector installation in STAR
		will allow for better particle identification from which charm and beauty
		searches will benefit considerably. In addition, the
		 future STAR Heavy Flavour Tracker \cite{HFT} under development is
		going to improve significantly the momentum resolution and the application
		of microvertexing techniques for charm and beauty identification and studies.

              \bibliographystyle{aipproc}   
              \bibliography{cipanp}
              \IfFileExists{\jobname.bbl}{}
                           {\typeout{}
                             \typeout{******************************************}
                             \typeout{** Please run "bibtex \jobname" to optain}
                             \typeout{** the bibliography and then re-run LaTeX}
                             \typeout{** twice to fix the references!}
                             \typeout{******************************************}
                             \typeout{}
                           }

              \end{document}